\documentclass[aps,pra,amsfonts,amssymb,amsmath,showpacs,
floatfix,nofootinbib,groupedaddress,superscriptaddress,citesort]{revtex4}
\usepackage{subfigure}
\usepackage{mathrsfs}
\usepackage{longtable}
\usepackage{amsfonts}
\usepackage{amstext}
\usepackage{amsmath}
\usepackage{amssymb}
\usepackage{amsthm}
\usepackage[usenames]{xcolor}
\usepackage[dvips]{graphicx}
\def\qed{\leavevmode\unskip\penalty9999 \hbox{}\nobreak\hfill
     \quad\hbox{\leavevmode  \hbox to.77778em{%
              \hfil\vrule   \vbox to.675em%
               {\hrule width.6em\vfil\hrule}\vrule\hfil}}
     \par\vskip3pt}

\newtheorem{*thm*}[lemma]{Theorem}

\begin{document}
\title{Complete Optimal Convex Approximations of Qubit States under $B_2$ Distance}
\author{Xiao-Bin Liang}
\affiliation{School of Mathematics and Computer science, Shangrao Normal University,
 Shangrao 334001, China}
\author{Bo Li}
\email{libobeijing2008@163.com}
\affiliation{School of Mathematics and Computer science, Shangrao Normal University,
 Shangrao 334001, China}\affiliation{Max-Planck-Institute
for Mathematics in the Sciences, 04103 Leipzig, Germany}
\author{Biao-Liang Ye}
\affiliation{Quantum Information Research Center,  Shangrao Normal University, Shangrao 334001, China}
\author{Shao-Ming Fei}
 \affiliation{Max-Planck-Institute
for Mathematics in the Sciences, 04103 Leipzig, Germany}\affiliation{School of Mathematical Sciences, Capital Normal University, Beijing 100048, China}
\author{Xianqing Li-Jost}
\affiliation{Max-Planck-Institute
for Mathematics in the Sciences, 04103 Leipzig, Germany}

\begin{abstract}
We consider the optimal approximation of arbitrary qubit states with respect to an available states
consisting the eigenstates of two of three Pauli matrices, the $B_2$-distance of an arbitrary target state. Both the analytical formulae of the $B_2$-distance and the corresponding complete
optimal decompositions are obtained. The tradeoff relations for both
the sum and the squared sum of the $B_2$-distances have been
analytically and numerically investigated.

\end{abstract}
\pacs{03.67.-a, 03.65.Ud,  03.65.Yz}
\maketitle

\section{Introduction}

Quantifying correlations among multipartite systems is one of the most important problems in quantum theory. However, most correlation measures become notorious
difficult to calculate with the increasing partite and dimension.
An alternative way  to deal with the problem is to consider
the distance of a given state to the so called free states in resource theory.
For example, entanglement is considered as the minimal
distance of a given state to the set of separable states in quantum systems \cite{horodecki1,wootters,enta,enta2}. The quantum discord is regarded as the minimal distance of a given
state to classically correlated states \cite{modi}. And quantum coherence can be quantified by the optimal convex approximation of the given state to the reference orthogonal base \cite{coh}.

While convexity is a very important property in mathematics and has been studied for long time, several related recent developments in quantum information have
stimulated new interest in this topic \cite{Jiangwei,Barreiro}.
The problem of optimal approximation to an unavailable quantum channel or state by the available channels or states was considered in \cite{ss,MFSacchi} recently.
It was shown that the optimally approximated distance has an natural operational interpretation. It can quantify the least distinguishable channel (state) from the given convex set to the target channel (state).  The trace distance measure of coherence can be regarded as convex approximation to the target state with respect to a fixed base of the system, where the fixed base can be either orthogonal or nonorthogonal\cite{coh3,enta4,cohq,tong,Theurer}.
In Ref. \cite{MFSacchi}, the author considered the the $B_3$-distance, the distance from a target qubit state to the convex approximation of bases containing the eigenstates of all Pauli matrices. The optimal convex approximation on the $B_3$-distance has been obtained.

In this work, we focus on $B_2$-distance, the distance corresponding to the convex approximation of bases containing the eigenstates from one of the pairs of Pauli matrices. We investigate all the optimal convex decompositions for the desired quantum state. The paper is organized as follows. In \ref{b2}, we calculate the $B_2$-distance in eight different cases, with the parameter regions achieving each optimal approximation explicitly given. In \ref{tradeoff}, we study tradeoff relations for both
the sum and the square sum of the $B_2$-distance.

\section{The Pauli $B_2-$distance of qubit state}\label{b2}

For an equal \emph{priori} probability of two given quantum states $\rho$ and $\rho_0$, the optimal discrimination between them can be quantified by the following probability $p_{discr}(\rho,\rho_0)$,
$$
p_{discr}(\rho,\rho_0)=1/2+1/4\parallel \rho-\rho_0\parallel_1,
$$
where $\parallel\rho\parallel_1$ denotes the trace norm of $\rho$,
$\parallel \rho\parallel_1=Tr \sqrt{\rho^\dag\rho}=\sum_i\sqrt{r_i}$,
$r_i$ are the eigenvalues of $\rho^\dag\rho$.

The optimal convex approximation of the quantum state $\rho$ with respect to a given set  $\rho_i$ is quantified by
$D_{\{\rho_i\}}(\rho)=\min_{\{p_i\}}\{\parallel\rho-\sum_i p_i\rho_i\parallel_1\}$, and the best  approximated points are the set $S(\rho^{opt})=\{\rho^{opt}|D_{\{\rho_i\}}(\rho)=\parallel\rho-\rho^{opt}\parallel_1\}$.

This optimal convex approximation provides the worst probability of discriminating the desired state $\rho$ from any of the available states $\sum_i p_i \rho_i$. For any other figure of merit that quantifies the distance between quantum states, the optimal convex approximation can be similarly defined (e.g., by a decreasing function of the fidelity). We remind that the best approximation can be arrived at many points and $S(\rho^{opt})$ represents the set of all the optimal points achieving the minimum distance.

Let $|0 \rangle$ and $|1\rangle$, $|2 \rangle \equiv \frac{1}{\sqrt 2} (|0 \rangle +|1 \rangle )$ and
$|3\rangle \equiv \frac {1}{\sqrt{2}}(|0 \rangle - |1 \rangle )$, and
$|4 \rangle \equiv \frac {1}{\sqrt{2}}(|0 \rangle + \sqrt{-1} |1 \rangle )$ and
$|5 \rangle \equiv \frac {1}{\sqrt{2}} (|0\rangle - \sqrt{-1} |1 \rangle )$
be the eigenstates of the Pauli matrices $\sigma_z$, $\sigma_x$, and $\sigma_y$, respectively.
We consider the following available set of states,
\begin{eqnarray}
{\sf B}_2^{'}  = \!\!\!\!\!&&
\left \{    |0 \rangle ,\, |1
\rangle , \,|2 \rangle \equiv \frac{1}{\sqrt 2} (|0 \rangle +|1 \rangle ),\, |3
\rangle \equiv \frac {1}{\sqrt 2}
(|0 \rangle - |1 \rangle )\right \},
\nonumber \\
{\sf B}_2^{''}  = \!\!\!\!\!&&
\left \{    |0 \rangle ,\, |1
\rangle , \,|4 \rangle \equiv \frac {1}{\sqrt 2}
(|0 \rangle + \sqrt{-1} |1 \rangle ),\, |5 \rangle \equiv \frac {1}{\sqrt 2} (|0
\rangle - \sqrt{-1} |1 \rangle )\right \},
\nonumber \\
{\sf B}_2^{'''}  = \!\!\!\!\!&& \left \{ |2 \rangle \equiv \frac{1}{\sqrt 2} (|0 \rangle +|1 \rangle ),\, |3
\rangle \equiv \frac {1}{\sqrt 2}
(|0 \rangle - |1 \rangle ), \,|4 \rangle \equiv \frac {1}{\sqrt 2}
(|0 \rangle + \sqrt{-1} |1 \rangle ), \,|5 \rangle \equiv \frac {1}{\sqrt 2} (|0
\rangle - \sqrt{-1} |1 \rangle )
\right \}
\;,\label{availabe123}
\end{eqnarray}
where ${\sf B}_2^{'}$ contains the eigenstates of $\sigma_x, \sigma_z$, ${\sf B}_2^{''}$ the eigenstates of $\sigma_y, \sigma_z$ and ${\sf B}_2^{'''}$ the eigenstates of $\sigma_x, \sigma_y$. The target qubit state $\rho $ can be parameterized by
\begin{eqnarray}
\rho =\left(
\begin{array}{cc} 1-a
& k
\sqrt{a(1-a)}e^{-\sqrt{-1}\phi }\\
k \sqrt{a(1-a)}e^{\sqrt{-1}\phi }
& a  \\ \end{array} \right ) \;,\label{rho123}
\end{eqnarray} with
$a \in [0,1]$, $\phi \in [0,2\pi]$, and $k\in [0, 1]$ \cite{MFSacchi}.
Since the ${\sf B}_2$-distance is invariant under the state transformations $
\rho (a,k,\phi)\rightarrow \rho (1-a,k,\phi)$ and
$\rho (a,k,n  \pi /2 \pm \phi)\rightarrow \rho (a,k,\phi)$ (with integer $n$),
we can restrict our study on the case $a \in [0,1/2]$ and $\phi \in [0,  \pi /2]$.

For any given target quantum state $\rho$ and available basis set in Eq.(\ref{availabe123}), we reduce the optimal approximation problem to find the minimum $D_{B_2}(\rho)=\min_{\{p_i\}} \{\Vert \rho - \textstyle \sum  p_i |e_i \rangle \langle e_i| \Vert_1\}$ with respect to the probabilities $\{ p_i \}$, where $|e_i\rangle$ represent the states of ${\sf B}_2^{'}$, ${\sf B}_2^{''}$ or ${\sf B}_2^{'''}$ in Eq.(\ref{availabe123}). Denote  $\rho_i=|i\rangle\langle i|$. The original problem is reduced to the optimal approximation problem  of finding the minimum distance
\begin{eqnarray*}
D_{B_2}(\rho)&=&\min_{\{p_i\}}\{2\sqrt{| Det(\rho-\sum _{i}  p_i\rho_i )|}\}
\end{eqnarray*}
such that $p_i\geq 0$, $\sum _{j}p_j=1$.

We first consider the ${\sf B}_2^{'}$ distance, i.e.,
$D_{B_2^{'}}(\rho)=\min_{\{p_i\}}\{2\sqrt{| Det(\rho-\sum_{i=0}^3  p_i\rho_i )|}\}$.
Set
\begin{eqnarray*}
f(p_0,p_1,p_2,p_3)=|Det(\rho-\sum_{i=0}^3p_i\rho_i)|-\sum_{i=0}^3\lambda_ip_i-\lambda\sum_{i=0}^3p_i.
\end{eqnarray*}
Since the  constraint  inequality condition sets $p_i\geq 0$ is convex and the equality constraint $\sum _{j}p_j=1$ is linear, by the Karush-Kuhn-Tucker Theorem\cite{Vandenberghe},
the following KKT condition  must be satisfied while solving the above optimization problem.
\begin{eqnarray}
\frac{\partial f}{\partial p_i} =0,~\lambda_ip_i=0,~\lambda_i\geq0,~p_i\geq0,~\sum^3 _{j=0}p_j=1,~i=0,1,2,3,\label{differientialoperator}
\end{eqnarray}
Eq.(\ref{differientialoperator}) reduces to the following equations,
\begin{eqnarray*}
&&p_1+\frac{1}{2}p_2+\frac{1}{2}p_3+\lambda_0+\lambda-a=0, \nonumber
\\& &
p_0+\frac{1}{2}p_1+\frac{1}{2}p_3+\lambda_1+\lambda-1+a=0,\nonumber \\& &
\frac{1}{2}p_0+\frac{1}{2}p_1+p_3+\lambda_2+\lambda+k\sqrt{a(1-a)}\cos\phi-\frac{1}{2}=0,\nonumber \\& &
\frac{1}{2}p_0+\frac{1}{2}p_1+p_2+\lambda_3+\lambda-k\sqrt{a(1-a)}\cos\phi-\frac{1}{2}=0,\nonumber \\& &
\lambda_i\,p_i=0,~\lambda_i\geq0,~p_i\geq0,~i=0,1,2,3,\nonumber \\& &
\Sigma _i\,p_i=1.
\end{eqnarray*}
Solving the above equations, we can obtain the complete analytical solutions to the optimal convex approximation $\rho^{opt}$ of $\rho$. The $S(\rho^{opt})$ of $\rho $ with
respect to ${\sf B}_2^{'}$ is given as

\noindent  $i)$
For $a\geq k{\sqrt{a(1-a)}}\cos\phi$, we have
\begin{eqnarray}
D_{{\sf B}_2^{'}}(\rho )= 2k \sqrt{a(1-a)}\sin\phi=\langle\sigma_y\rangle,
\;\label{dunob1}
\end{eqnarray}
which is attained at
\begin{eqnarray}
&&p_0=1-a-k\sqrt{a(1-a)}\cos \phi -t,\nonumber
\\& &
p_1=a-k\sqrt {a(1-a)}\cos \phi-t,\nonumber \\& &
p_2=2k\sqrt {a(1-a)}\cos \phi+t,\nonumber \\& &
p_3=t,\label{pi11}
\end{eqnarray}
where $t$ satisfies $a-k{\sqrt{a(1-a)}}\cos\phi\geq t\geq0$.
Let $ A_1^{'}=\{\Sigma^3_{i=0} p_i\rho_i \}$ denote the set of states with $p_i$ given by Eq.(\ref{pi11}).
Then $A_1^{'}$ contains all the optimal points achieving the distance $D_{{\sf B}_2^{'}}(\rho )$ in Eq.(\ref{dunob1}).

\par \noindent  $ii)$
For  $a< k{\sqrt{a(1-a)}}\cos\phi$, we have the
the optimal convex approximation distance
\begin{eqnarray}
D_{{\sf B}_2^{'}}(\rho )&=& \sqrt{2(1+\sin^2\phi)k^2(a(1-a))-4a\cos\phi k\sqrt{a(1-a)}+2a^2}\nonumber \\
&=&\sqrt{\langle\sigma_y\rangle^2+\frac{1}{2}(\langle\sigma_x\rangle+\langle\sigma_z\rangle-1)^2},\label{dunobb2}
\end{eqnarray}
which is attained with
\begin{eqnarray}
&&p_0=1-a-k\sqrt{a(1-a)}\cos \phi, \nonumber
\\& &
p_2=a+k\sqrt {a(1-a)}\cos \phi,\nonumber \\& &
p_1=p_3=0.\label{PI22}
\end{eqnarray}
Denote $ A_2^{'}=\{p_0\rho_0+ p_2\rho_2\}$, with $p_0,~p2$ given by Eq.(\ref{PI22}). Then $ A_2^{'}$ contains all the optimal states achieving the distance
$D_{{\sf B}_2^{'}}(\rho )$ in Eq.(\ref{dunobb2}). Therefore $S(\rho^{opt})$ is given by
$S(\rho^{opt})=A_1^{'}\bigcup A_2^{'}$, which is the set of optimal states that gives rise to the
optimal convex approximations.

Next we consider the optimal convex approximation of $\rho $ with
respect to ${\sf B}_2^{''}$. Namely,
$D_{B_2^{''}}(\rho)=\min_{\{p_i\}}\{2\sqrt{| Det(\rho-\sum_{i=0,1,4,5}  p_i\rho_i )|}\}$.
Similar to the case of ${\sf B}_2^{'}$, we have

\par \noindent  $i)$
For $a\geq k {\sqrt{a(1-a)}}\sin\phi$, the optimal convex approximated distance is given by
\begin{eqnarray}
D_{{\sf B}_2^{''}}(\rho )= 2k \sqrt{a(1-a)}\cos\phi=\langle\sigma_x\rangle.
\label{dunob214}
\end{eqnarray}
The with the optimal probability weights are given by
\begin{eqnarray}
&&p_0=1-a-k\sqrt{a(1-a)}\sin \phi-t, \nonumber
\\& &
p_1=a-k\sqrt {a(1-a)}\sin \phi\nonumber-t, \\& &
p_4=2k\sqrt {a(1-a)}\sin \phi\nonumber +t,\\& &
p_5=t,
\label{pbii2}
\end{eqnarray}
where $t$ satisfies $a-k{\sqrt{a(1-a)}}\sin\phi\geq t\geq0$.
Denote $ A_1^{''}=\{\Sigma p_i\rho_i \}$ with $p_i$ given by Eq.(\ref{pbii2}).
Then $A_1^{''}$ contains all the optimal states achieving the distance $D_{{\sf B}_2^{''}}(\rho )$ in Eq.(\ref{dunob214}).

\par \noindent  $ii)$
For $a<k {\sqrt{a(1-a)}}\sin\phi $, we have the optimal convex approximated distance
\begin{eqnarray}
D_{{\sf B}_2^{''}}(\rho )&=& \sqrt{2(1+\cos^2\phi)k^2(a(1-a))-4a\sin\phi k\sqrt{a(1-a)}+2a^2}\nonumber \\
&=&\sqrt{\langle\sigma_x\rangle^2+\frac{1}{2}(\langle\sigma_y\rangle+\langle\sigma_z\rangle-1)^2},\label{duno133}
\end{eqnarray}
with the optimal probability weights given by
\begin{eqnarray}
&&p_0=1-a-k\sqrt{a(1-a)}\sin \phi, \nonumber
\\& &
p_4=a+k\sqrt {a(1-a)}\sin \phi,\nonumber \\& &
p_1=p_5=0.
\label{pbii23}
\end{eqnarray}
Let $ A_2^{''}=\{p_0\rho_0+ p_4\rho_4\}$ be the set of states with $p_0$ and $p_4$ given by Eq.(\ref{pbii23}). Then $ S(\rho^{opt})$ is given by $ S(\rho^{opt})=A_1^{''}\bigcup A_2^{''}$.

For the optimal approximation of $\rho $ with
respect to the basis in ${\sf B}_2^{'''}$, we have

\par \noindent  $i)$
For $1/2\geq k\sqrt{a(1-a)}(\sin\phi+\cos\phi)$, the optimal convex approximated distance has the form
\begin{eqnarray}
D_{{\sf B}_2^{'''}}(\rho )= (1-2a)=\langle\sigma_z\rangle,\label{dunoc3}
\end{eqnarray}
with the optimal probability weights
\begin{eqnarray}
&&p_2=1/2+k\sqrt{a(1-a)}(\cos\phi-\sin \phi)-t, \nonumber
\\& &
p_3=1/2-k\sqrt{a(1-a)}(\cos\phi+\sin \phi)-t, \nonumber  \\& &
p_4=2k\sqrt{a(1-a)}\sin \phi+t, \nonumber  \\& &
p_5=t,\label{dunoc3P3} \end{eqnarray}
where $t$ is given by $1/2\geq k\sqrt{a(1-a)}(\sin\phi+\cos\phi)\geq t\geq0$. Hence
$A_1^{'''}=\{\Sigma p_i\rho_i \}$, with $p_i$ given by Eq.(\ref{dunoc3P3}),
contains all the optimal states achieving
the distance $D_{{\sf B}_2^{'''}}(\rho )$ in Eq.(\ref{dunoc3}).

\par \noindent  $ii)$
For  $\frac {1}{2}<
k\sqrt{a(1-a)}(\sin\phi+\cos\phi)$, we have
\begin{eqnarray}
D_{{\sf B}_2^{'''}}(\rho )&=& \sqrt{(1-2a)^2+2(k \sqrt{a(1-a)}(\cos\phi+\sin\phi)-1/2)^2}\nonumber \\
&=&\sqrt{\langle\sigma_z\rangle^2+\frac{1}{2}(\langle\sigma_y\rangle+\langle\sigma_x\rangle-1)^2},\label{dunoa37}
\end{eqnarray}
with the optimal probability weights given by
\begin{eqnarray}
&&p_2=1/2+k\sqrt{a(1-a)}(\cos\phi-\sin \phi),\nonumber
\\& &
p_4=1/2-k\sqrt{a(1-a)}(\cos\phi-\sin \phi),\nonumber \\& &
p_3=p_5=0.\label{dunoa3p34}
\end{eqnarray}
Denoting $ A_2^{'''}=\{p_2\rho_2+ p_4\rho_4\}$, with $p_2$ and $p_4$ given by Eq.(\ref{dunoa3p34}), we have $S(\rho^{opt})=A_1^{'''}\bigcup A_2^{'''}$, which is the set of states achieving all the optimal convex approximations.

\begin{figure*}[htp]
\centering
\subfigure[$\phi=\pi/4$] {\includegraphics[height=2in,width=2in,angle=0]{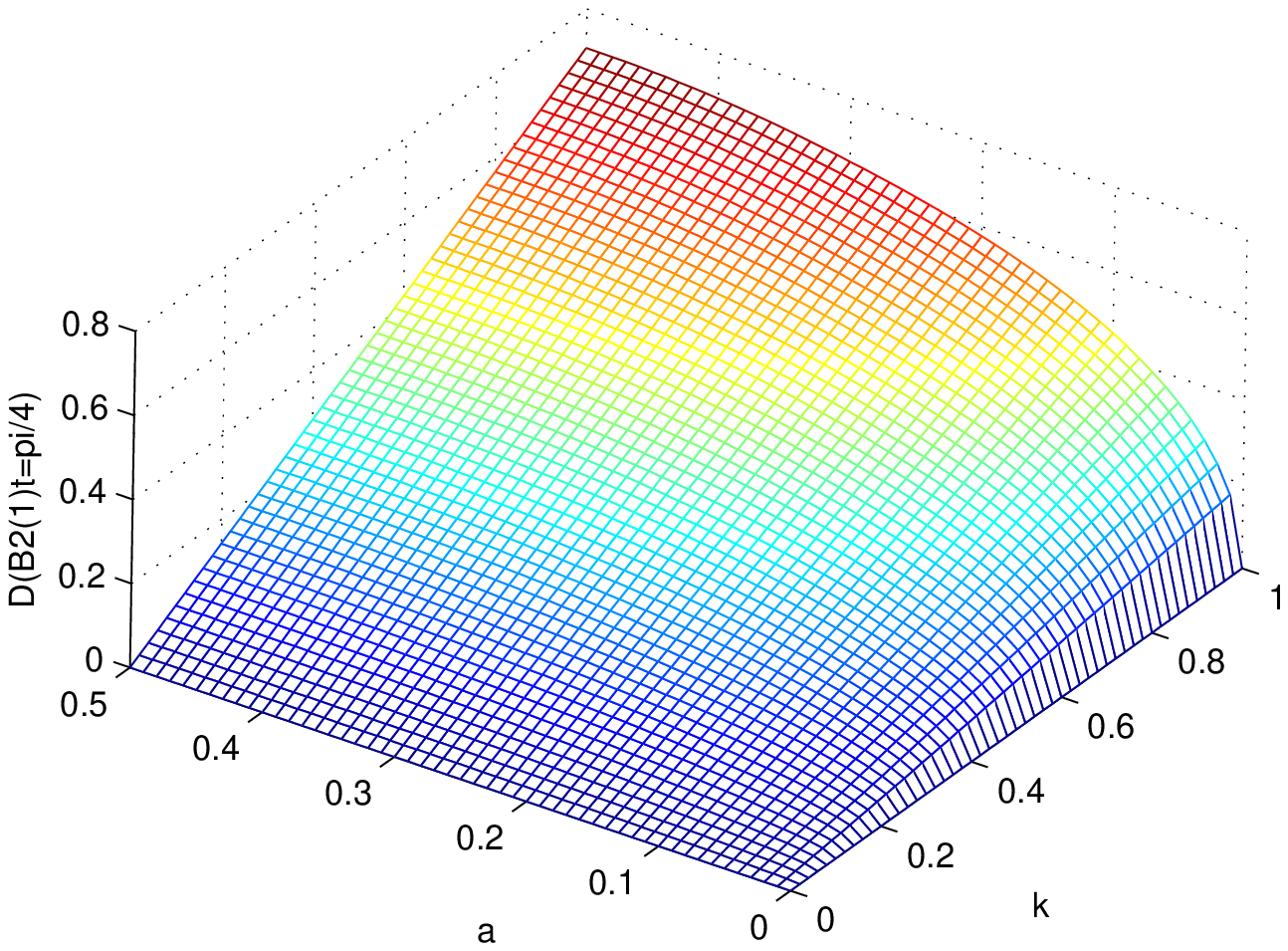}}
\subfigure[$k=4/5$] {\includegraphics[height=2in,width=2in,angle=0]{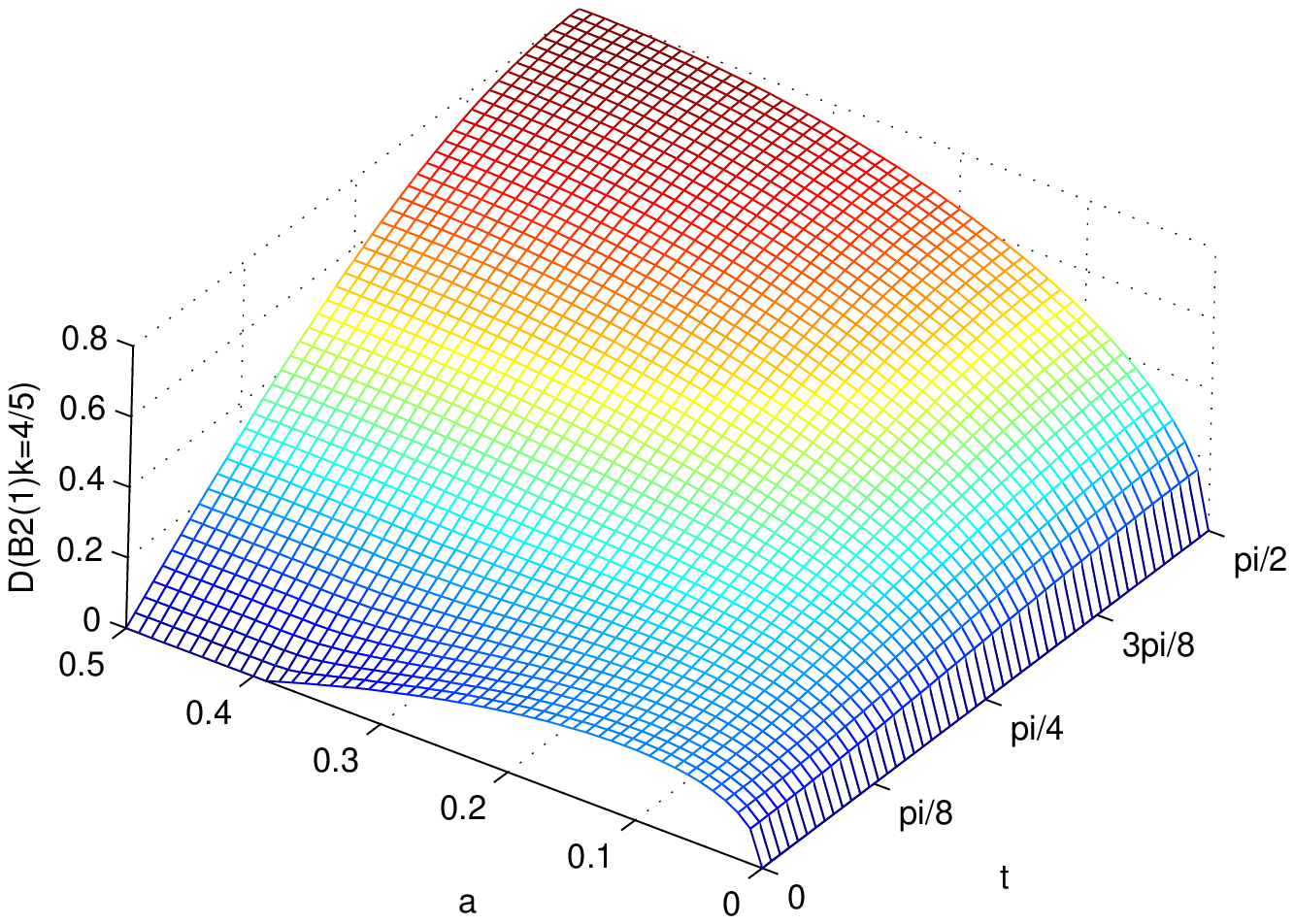}}
\subfigure[Area of the interface] {\includegraphics[height=2in,width=2in,angle=0]{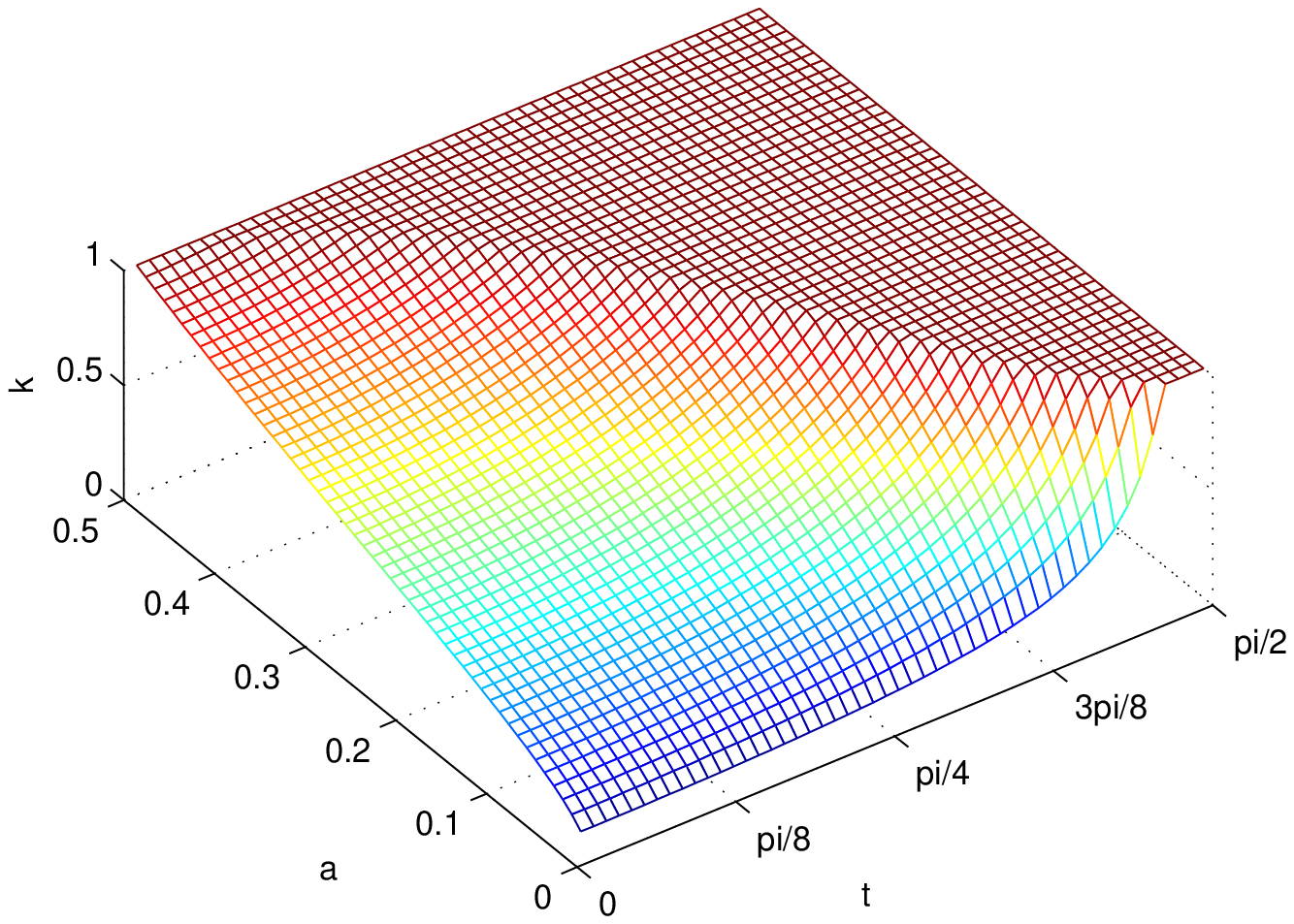}}
\caption{Optimal convex approximation of a qubit mixed state $\rho$
w.r.t. the set $B^{'}_2$ spanned by the eigenstates of the Pauli matrices $\sigma_z$ and $\sigma_x$.
The distance $D_{B^{'}_2}(\rho)$ is plotted vs the target state parameters $a,~k$ and $\phi$, for fixed value of the parameter $\phi= \frac{\pi}{4}$ [FIG. 1.(a)], for fixed value of the parameter $k = 4/5$ [FIG. 1.(b)]. The interface of the regions of the two cases i) and ii) is plotted in FIG. 1.(c), the region above the surface corresponds to the case $a\geq k{\sqrt{a(1-a)}}\cos\phi$, the region below the surface is the case $a< k{\sqrt{a(1-a)}}\cos\phi$.}
\label{fig1}
\end{figure*}

\begin{figure*}[htp]
\centering
\subfigure[$\phi=\pi/4$] {\includegraphics[height=2in,width=2in,angle=0]{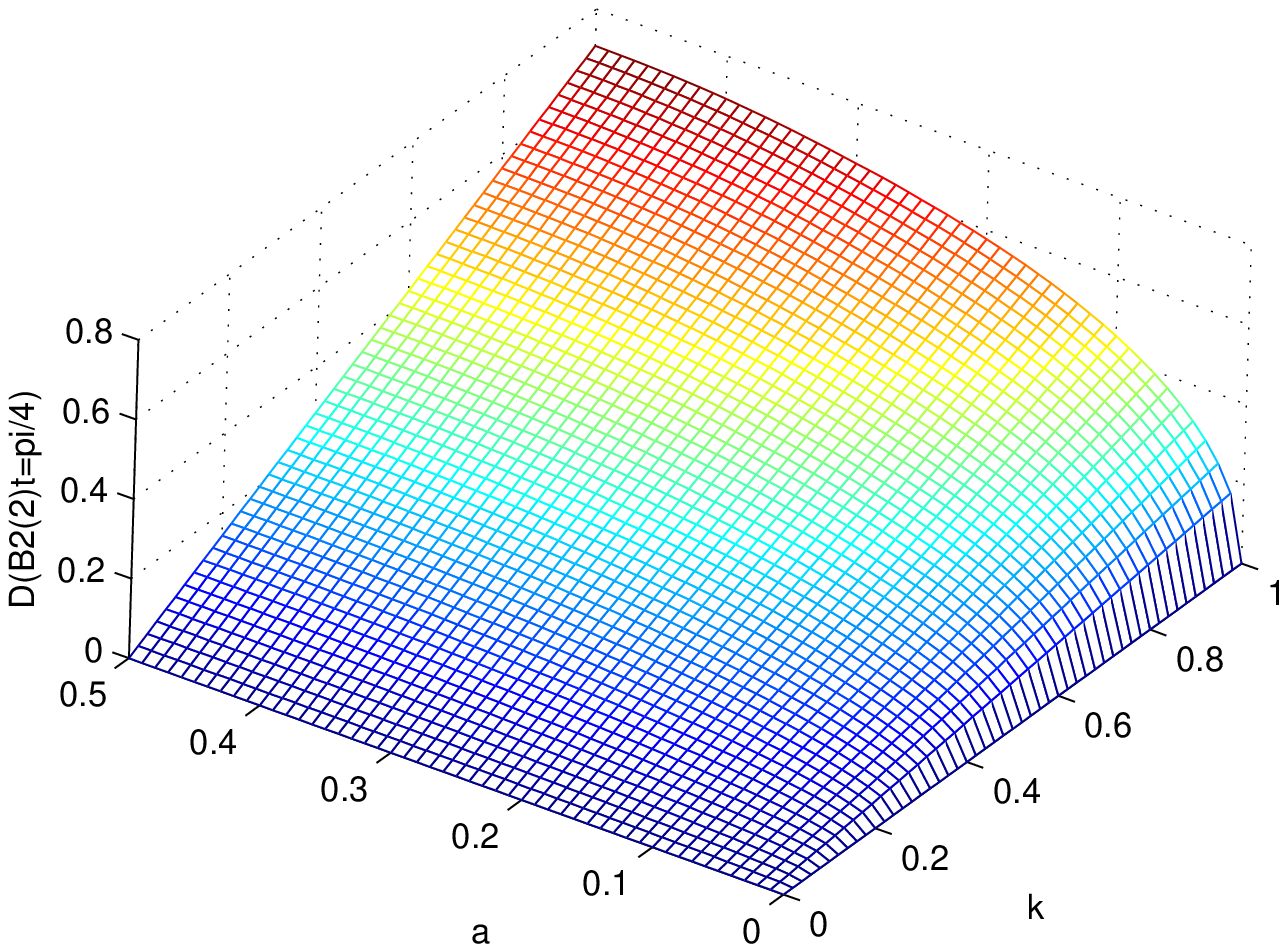}}
\subfigure[$k=4/5$] {\includegraphics[height=2in,width=2in,angle=0]{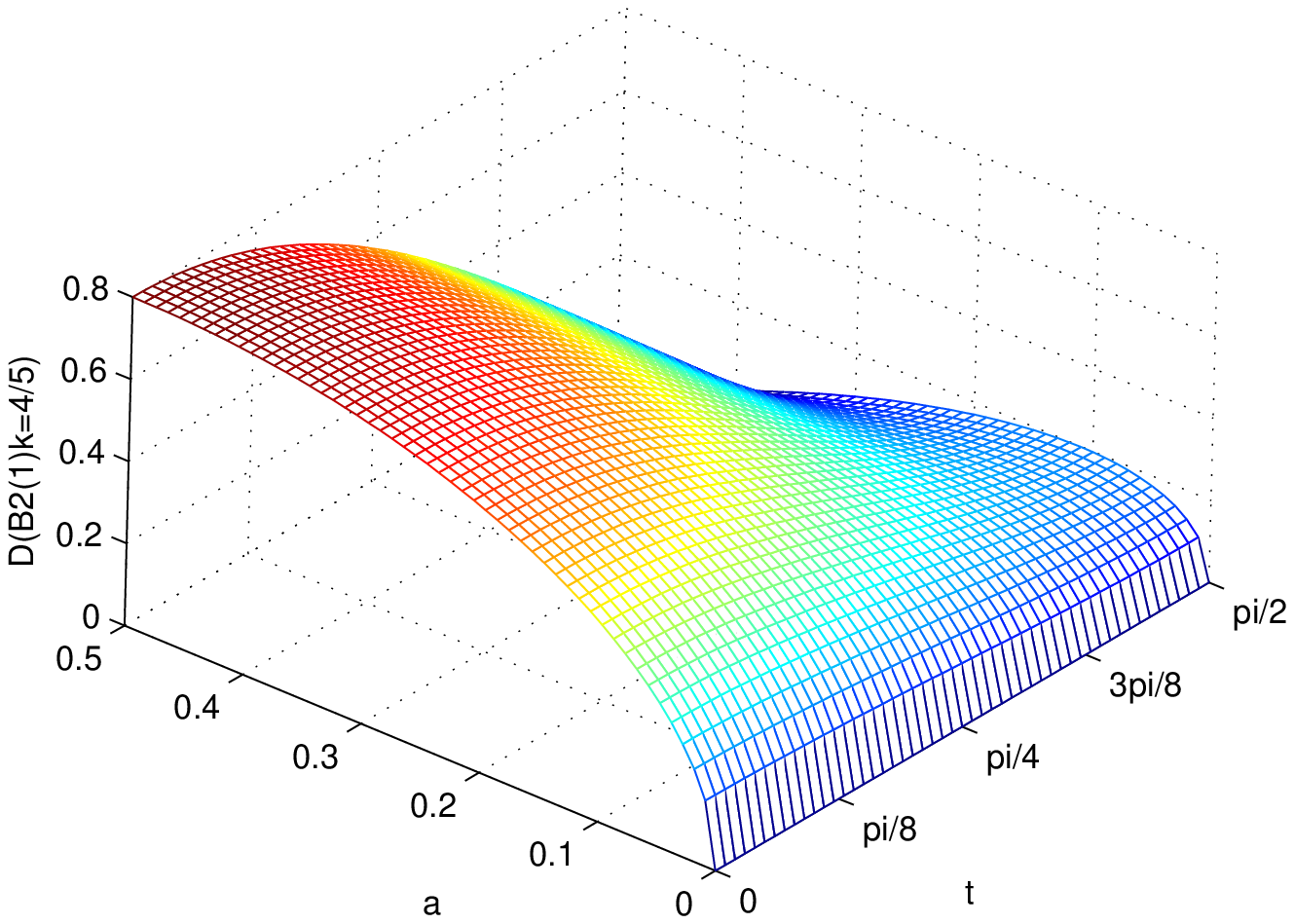}}
\subfigure[Area of the interface] {\includegraphics[height=2in,width=2in,angle=0]{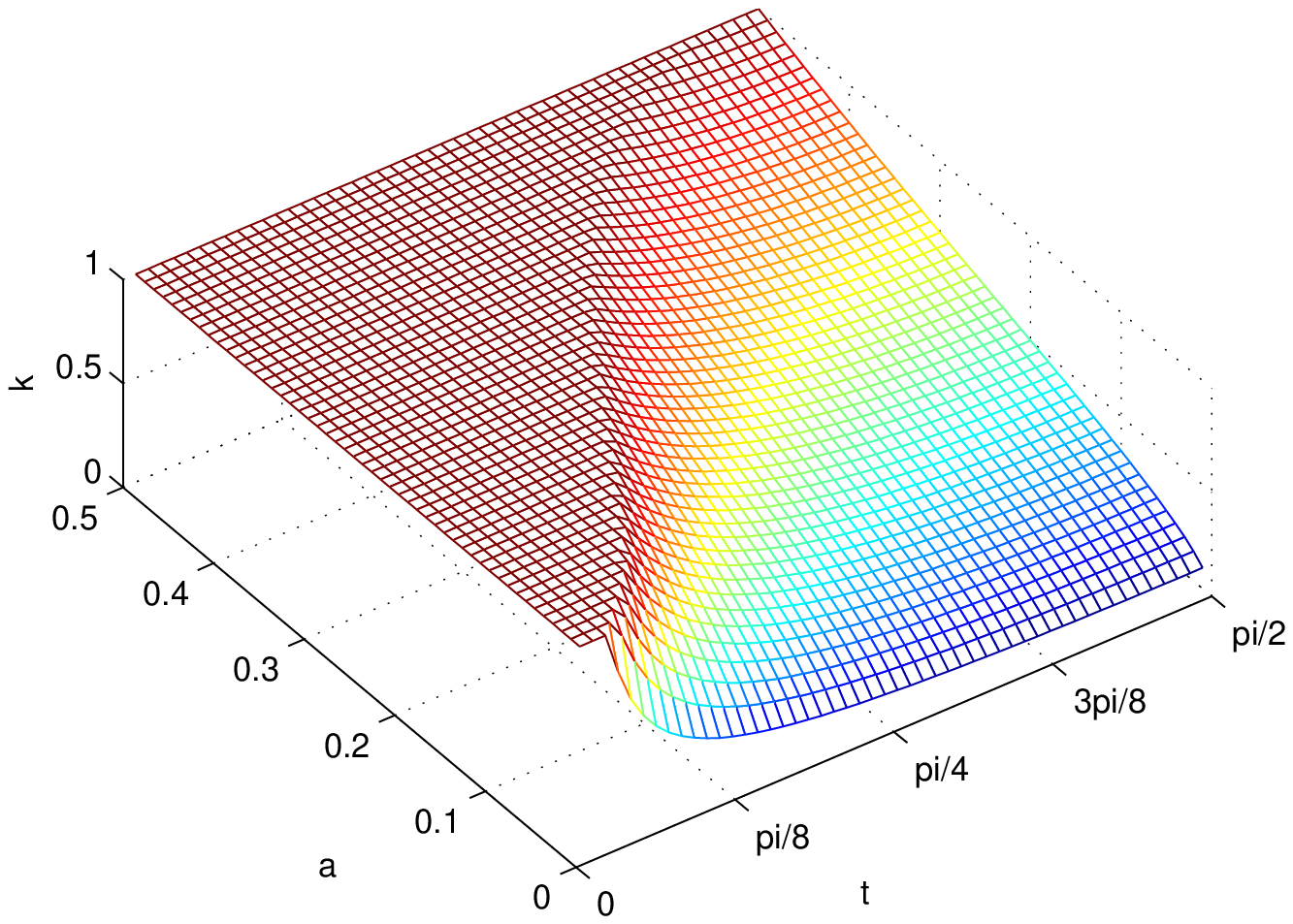}}
\caption{Optimal convex approximation of a qubit mixed state $\rho$
w.r.t. the set $B^{''}_2$ spanned by the eigenstates of the Pauli matrices $\sigma_z$ and $\sigma_y$. The
distance $D_{B^{''}_2}(\rho)$ is plotted for fixed value of the parameter $\phi= \frac{\pi}{4}$ [FIG.2.(a)], and for fixed value of the parameter $k = 4/5$ [FIG. 2.(b)]. The interface of the regions of the two cases i) and ii) is plotted in FIG. 2.(c), the region above the surface corresponds to the case  $a\geq k {\sqrt{a(1-a)}}\sin\phi$, the region below the surface is the case $a< k{\sqrt{a(1-a)}}\sin\phi$.
 }
\label{fig2}
\end{figure*}

\begin{figure*}[htp]
\centering
\subfigure[$\phi=\pi/4$] {\includegraphics[height=2in,width=2in,angle=0]{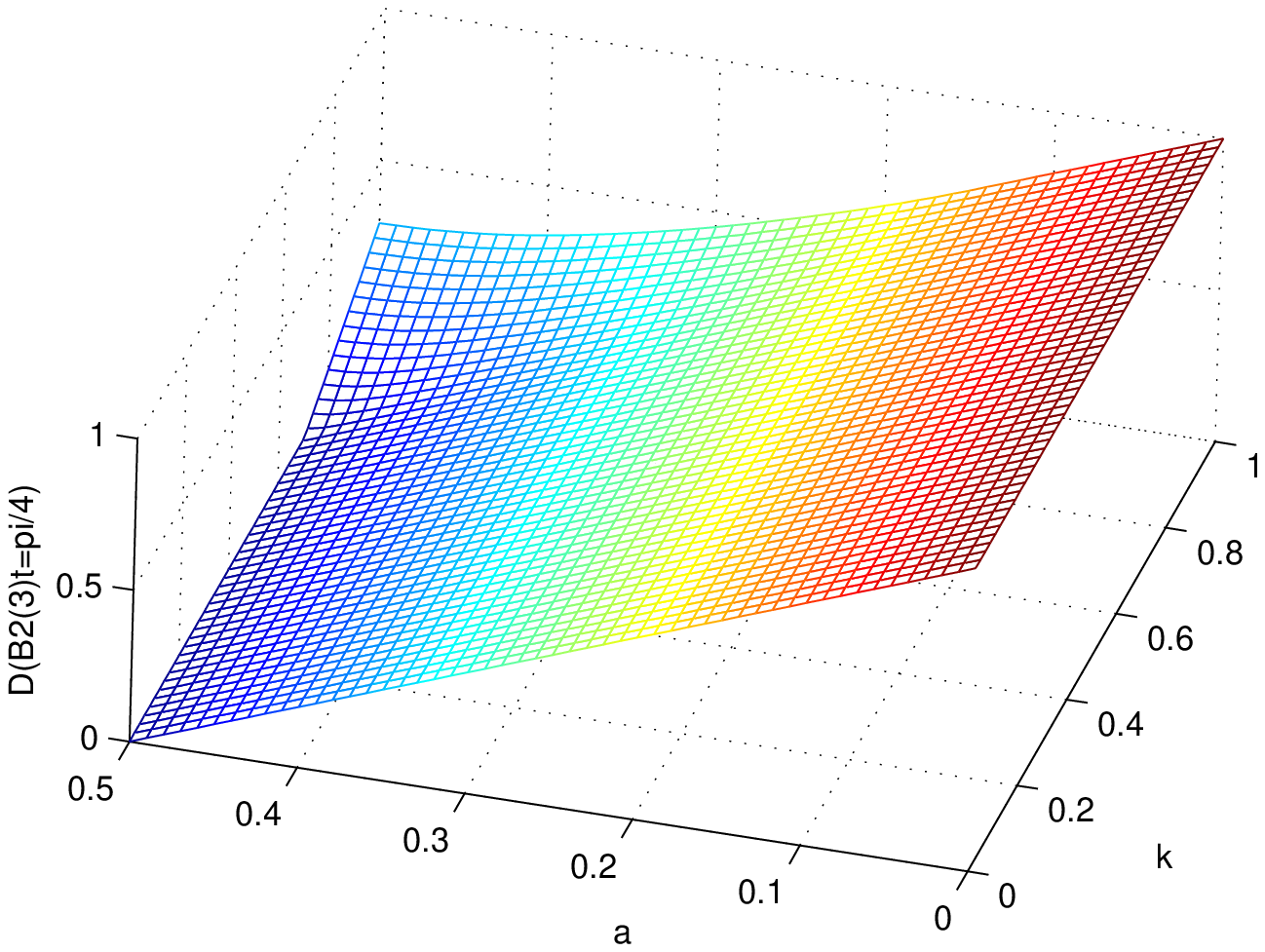}}
\subfigure[$k=4/5$] {\includegraphics[height=2in,width=2in,angle=0]{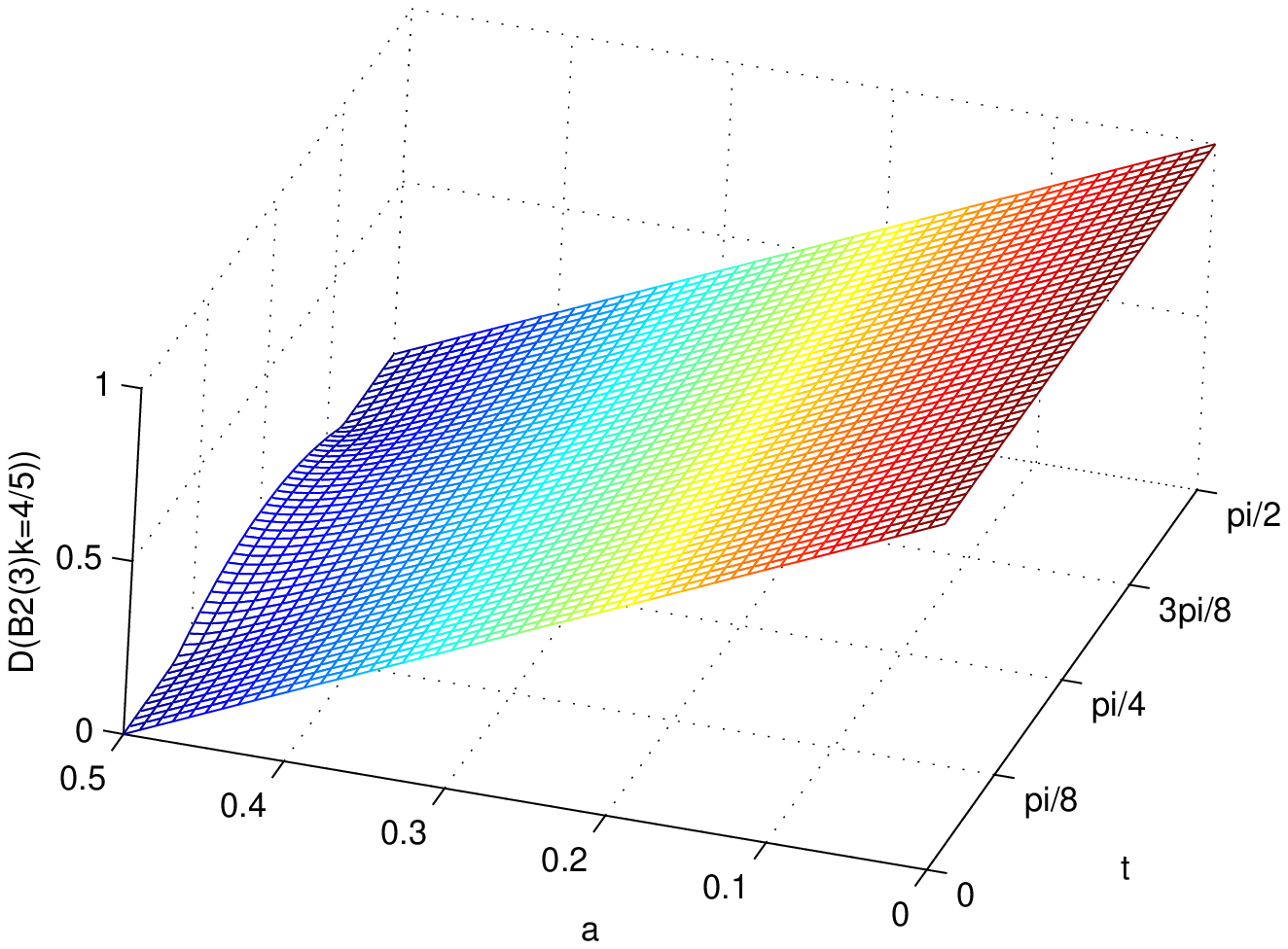}}
\subfigure[Area of the interface] {\includegraphics[height=2in,width=2in,angle=0]{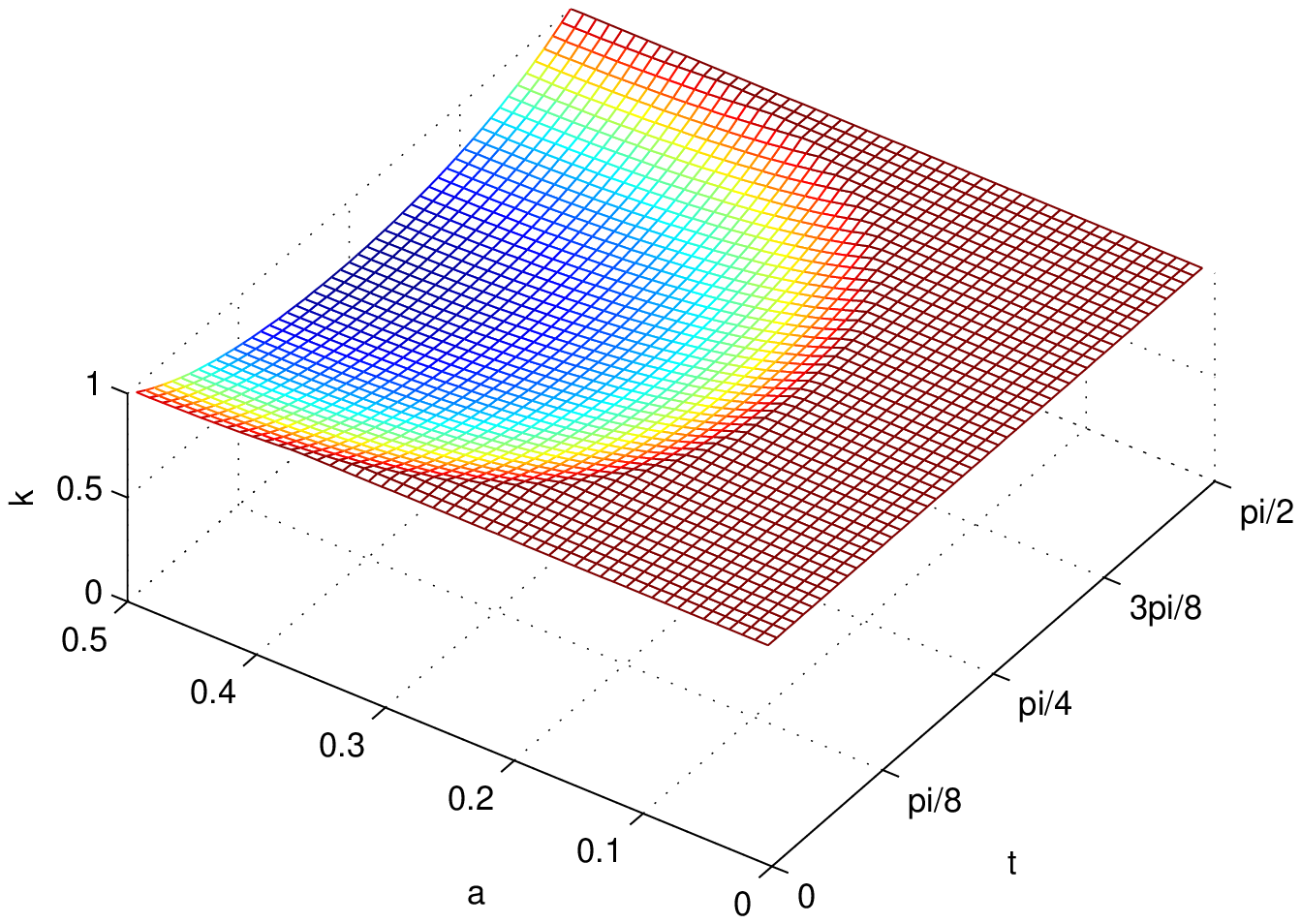}}
\caption{Optimal convex approximation of a qubit mixed state $\rho$
w.r.t. the set $B^{'''}_2$ spanned by the eigenstates of the Pauli matrices $\sigma_x$ and $\sigma_y$.
The distance $D_{B^{'''}_2}(\rho)$ is plotted for fixed value of the parameter $\phi= \frac{\pi}{4}$ [FIG.3.(a)] and for fixed value of the parameter $k = 4/5$ [FIG. 3.(b)]. The interface of the regions
of the two cases i) and ii) is plotted in FIG. 3.(c), the region above the surface corresponds to the case  $1/2\geq k\sqrt{a(1-a)}(\sin\phi+\cos\phi)$, the region below the surface is the case  $1/2< k\sqrt{a(1-a)}(\sin\phi+\cos\phi)$. }
\label{fig3}
\end{figure*}

In Fig. \ref{fig1}, we plot the distance $D_{{\sf B}_2^{'}}(\rho )$ for fixed parameters of $k$ and $\phi$. One can see that for the fixed value $\phi=\frac{\pi}{4}$, Fig.\ref{fig1}(a) shows that
the optimal distance $D_{{\sf B}_2^{'}}(\rho )$ increases with $k$ and decreases with the parameter $a$.
Fig.\ref{fig1}(c) shows the interface such that
the region above the surface corresponds to the case
$i)$, namely, $a\geq k{\sqrt{a(1-a)}}\cos\phi$; and the region below the surface is the case $ii)$, $a< k{\sqrt{a(1-a)}}\cos\phi$.
In Fig. \ref{fig2} and \ref{fig3}, the distances $D_{{\sf B}_2^{''}}(\rho )$ and $D_{{\sf B}_2^{'''}}(\rho )$ with the fixed values are also plotted, respectively.
The corresponding interface is plotted in Fig.\ref{fig2}(c) (Fig.\ref{fig3}(c)): the region above the surface corresponds to the case $a\geq k {\sqrt{a(1-a)}}\sin\phi$
($1/2\geq k\sqrt{a(1-a)}(\sin\phi+\cos\phi)$), the region below the surface is the case
$a< k {\sqrt{a(1-a)}}\sin\phi$ ($1/2< k\sqrt{a(1-a)}(\sin\phi+\cos\phi)$), respectively.

\section{Tradeoff relations among the $B_2-$distances}\label{tradeoff}
We have calculated the optimal distances $D_{B^{'}_2}(\rho)$, $D_{B^{''}_2}(\rho)$ and $D_{B^{'''}_2}(\rho)$, with explicit formulae for arbitrary qubit-state $\rho$ classified in
two parameter regions each. Interestingly, we find that
the sum and the squared sum of $D_{B^{'}_2}(\rho)$, $D_{B^{''}_2}(\rho)$ and $D_{B^{'''}_2}(\rho)$ display some tradeoff relations in each parameter region.

Let \textcircled{1} represents the parameter region of the state $\rho$ with $a\geq k{\sqrt{a(1-a)}}\cos\phi\}$, and \textcircled{2} the parameter region $a< k{\sqrt{a(1-a)}}\cos\phi\}$.
Similarly, \textcircled{3} (\textcircled{4}) represents the parameter region with $a\geq k{\sqrt{a(1-a)}}\sin\phi\}$ ($a< k{\sqrt{a(1-a)}}\sin\phi\}$), and
\textcircled{5} (\textcircled{6}) represents the parameter region
with $\frac{1}{2}\geq k{\sqrt{a(1-a)}}\cos\phi$ ($\frac{1}{2}<k{\sqrt{a(1-a)}}\cos\phi$).
For every $D_{B_2^{'}} (\rho)$ ($D_{B_2^{''}}(\rho)$, $D_{B_2^{'''}}(\rho)$) there are two
parameter regions: i) and ii). A state $\rho$ may belong to the region \textcircled{1}
in calculating the distance $D_{B^{'}_2}(\rho)$, but to region \textcircled{4} in calculating $D_{B^{''}_2}(\rho)$, and to region \textcircled{5} in calculating $D_{B^{'''}_2}(\rho)$.
Therefore we have all together eight cases of parameter regions

\begin{tabular}{|c|c|c|c|c|c|c|c|}
\hline
 region 1 &\textcircled{1}$\cap$\textcircled{3}$\cap$\textcircled{5} & region 3 &\textcircled{1}$\cap$\textcircled{3}$\cap$\textcircled{6} &region 5 &\textcircled{2}$\cap$\textcircled{3}$\cap$\textcircled{5}&region 7 &\textcircled{1}$\cap$\textcircled{4}$\cap$\textcircled{6}\\
 \hline
 region 2 &\textcircled{2}$\cap$\textcircled{4}$\cap$\textcircled{5}& region 4  & \textcircled{1}$\cap$\textcircled{4}$\cap$\textcircled{5} &region 6 &\textcircled{2}$\cap$\textcircled{3}$\cap$\textcircled{6}&region 8 &\textcircled{2}$\cap$\textcircled{4}$\cap$\textcircled{6}\\
\hline
\end{tabular}

\begin{figure*}[htp]
\centering
\subfigure[$\phi=\pi/4$, $\min D_{B_{2}}(\rho)$] {\includegraphics[height=2in,width=2in,angle=0]{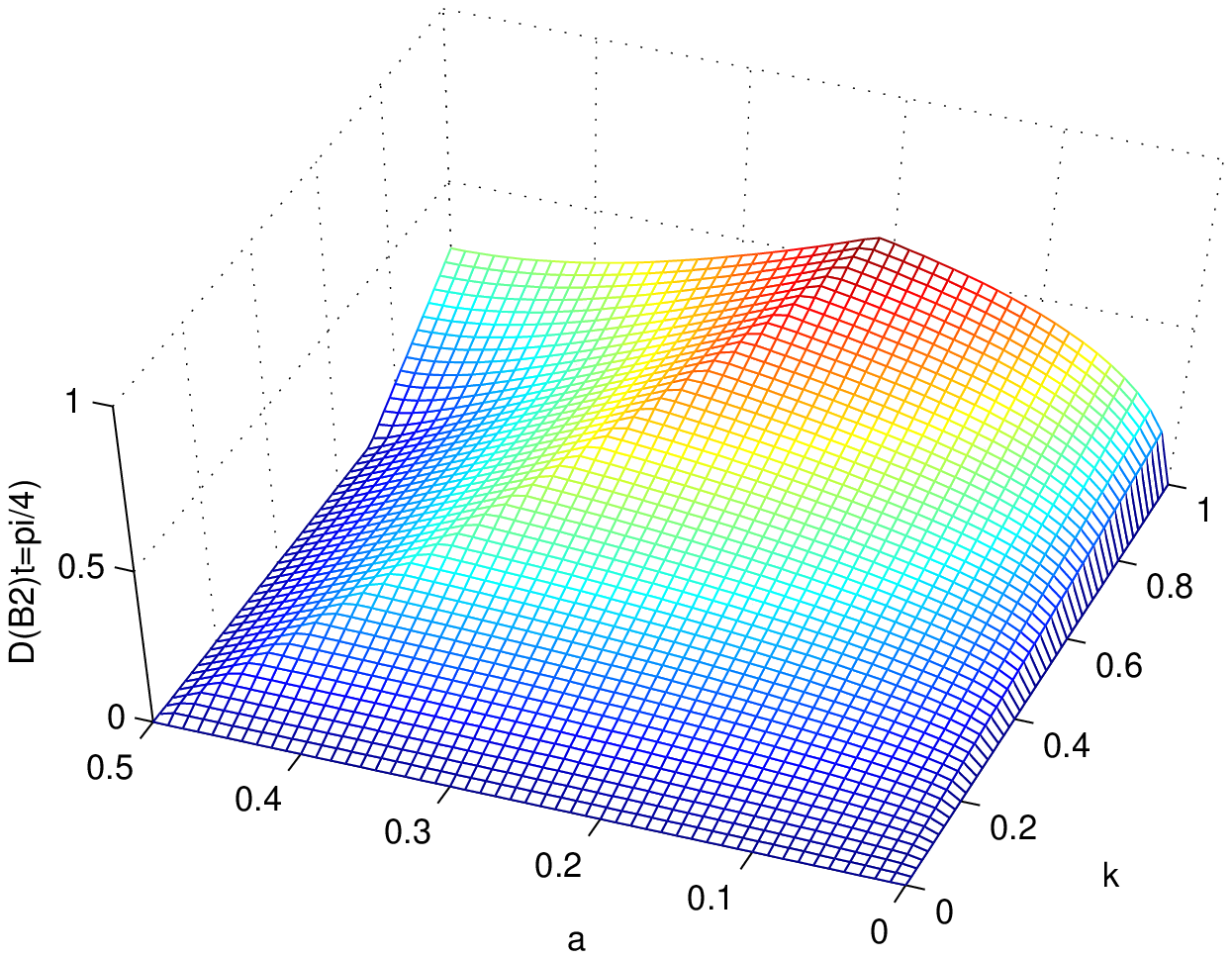}}
\subfigure[$k=4/5$, $\min D_{B_{2}}(\rho)$] {\includegraphics[height=2in,width=2in,angle=0]{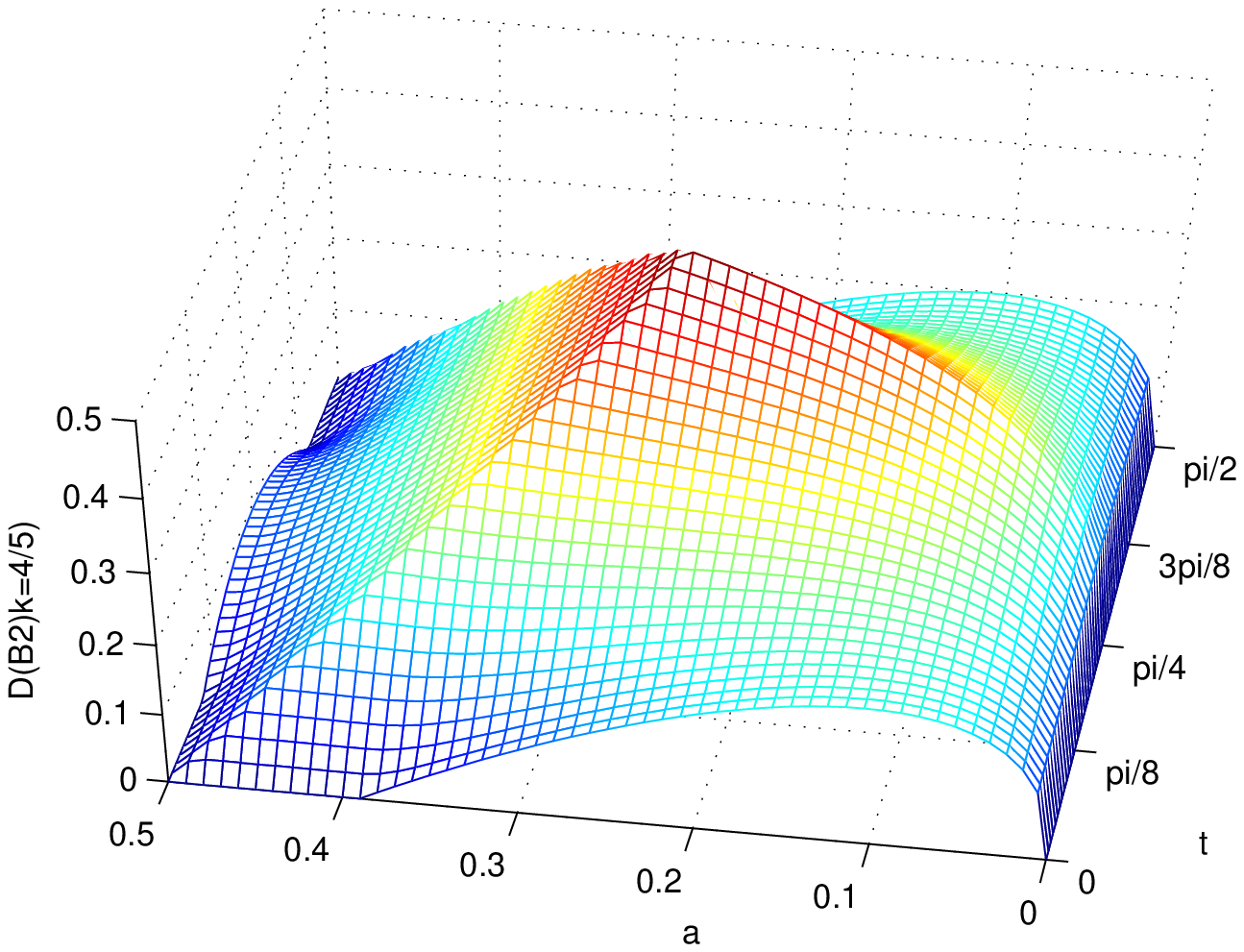}}
\caption{Illustrations of the minimum  Pauli $B_{2}-$distance, $\min D_{B_{2}}(\rho)=\min\{D_{B^{'}_{2}(\rho)},D_{B^{''}_{2}(\rho)},D_{B^{'''}_{2}(\rho)}$\} }
\label{fig4}
\end{figure*}

For each case, the three $B_2-$distances display different values. In Fig.\ref{fig4}, we plot the minimum  $\min D_{B_2}(\rho)=\min\{D_{B_2^{'}}(\rho),D_{B_2^{''}}(\rho),D_{B_2^{'''}}(\rho)\}$. One can see from Fig.\ref{fig4} that, for fixed $\phi=\pi/4$ and
$k=4/5$, $\min D_{B_2}(\rho)$ is always nonzero for nonzero parameters $a$ and $k$ or $\phi$, namely,
all the three Pauli $B_2-$distances are nonzero.

Concerning the tradeoff relations of the three Pauli $B_2-$distances, for convenience, we denote
\begin{eqnarray}
D_{B_2}(\rho)&=&D_{B_2^{'}}(\rho)+D_{B_2^{''}}(\rho)+D_{B_2^{'''}}(\rho),\\
D_{B_2}(\rho)^2&=&D_{B_2^{'}}(\rho)^2+D_{B_2^{''}}(\rho)^2+D_{B_2^{'''}}(\rho)^2.
\label{DB22}
\end{eqnarray}
By the numerical calculation, we obtain the tradeoff relation among $D_{B_2^{'}}(\rho)$, $D_{B_2^{''}}(\rho)$ and $D_{B_2^{'''}}(\rho)$, see the following table:

\begin{tabular}{|c|c|c|c|c|c|}
\hline
region &$D_{B_2}(\rho)$&$D_{B_2}(\rho)^2$&region&$D_{B_2}(\rho)$&$D_{B_2}(\rho)^2$\\
\hline
1&[0,1.742)&[0.1]&5&[1,1.742)&(0.501,1.086)\\
\hline
2&(1.006,1.750)&(0.666,1.068)&6&(1,1.742)&(0.666,1.068)\\
\hline
3&[1,1.742)&(0.501,1.086)&7&(1,1.742)&(0.666,1.068)\\
\hline
4&[1,1.742)&(0.501,1.086)&8&(1.5,1.765)&(0.750,1.060)\\
\hline
\end{tabular}

\begin{figure*}[htp]
\centering
\subfigure[Over  view ] {\includegraphics[height=2in,width=2in,angle=0]{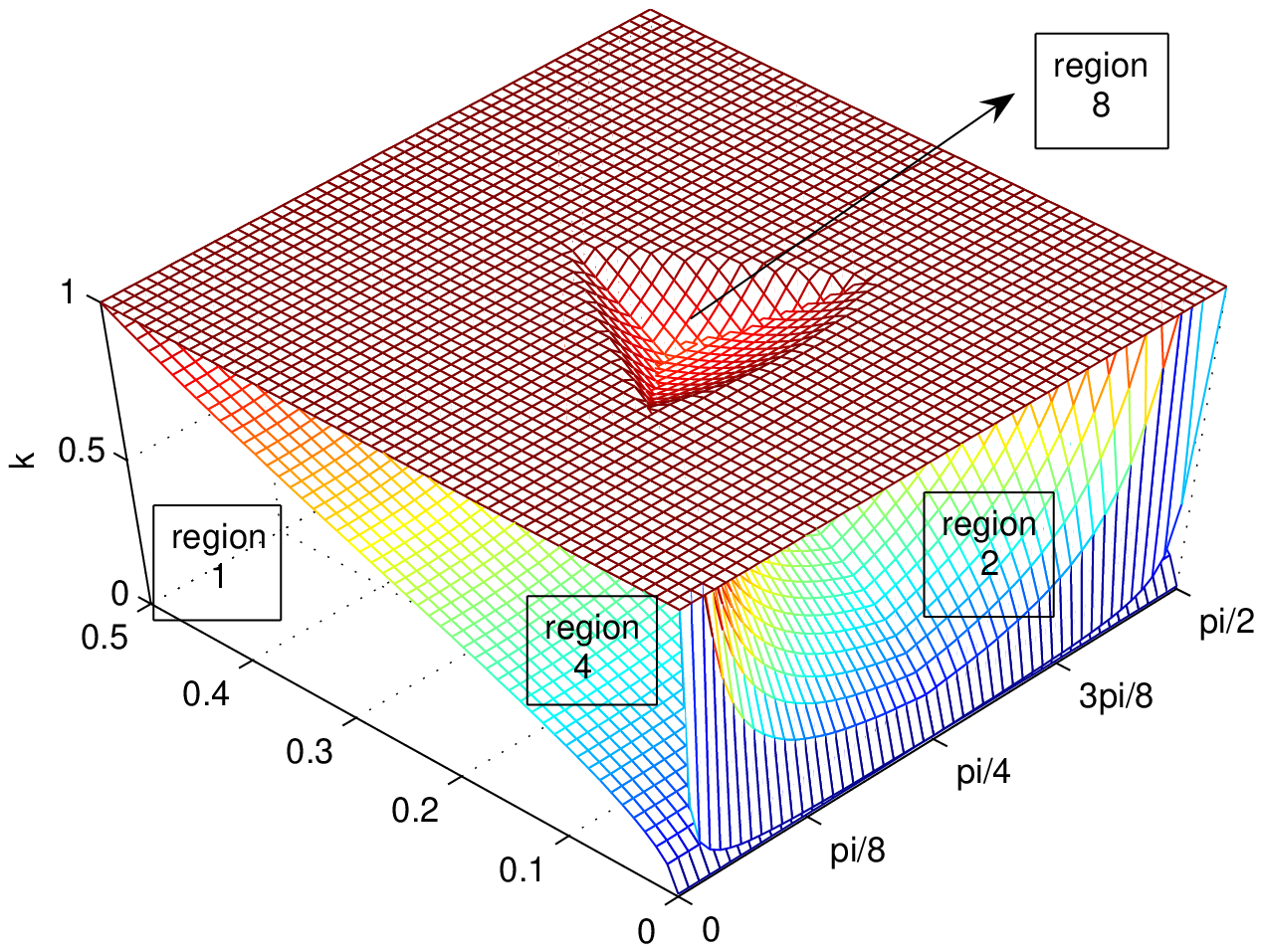}}
\subfigure[upward view ] {\includegraphics[height=2in,width=2in,angle=0]{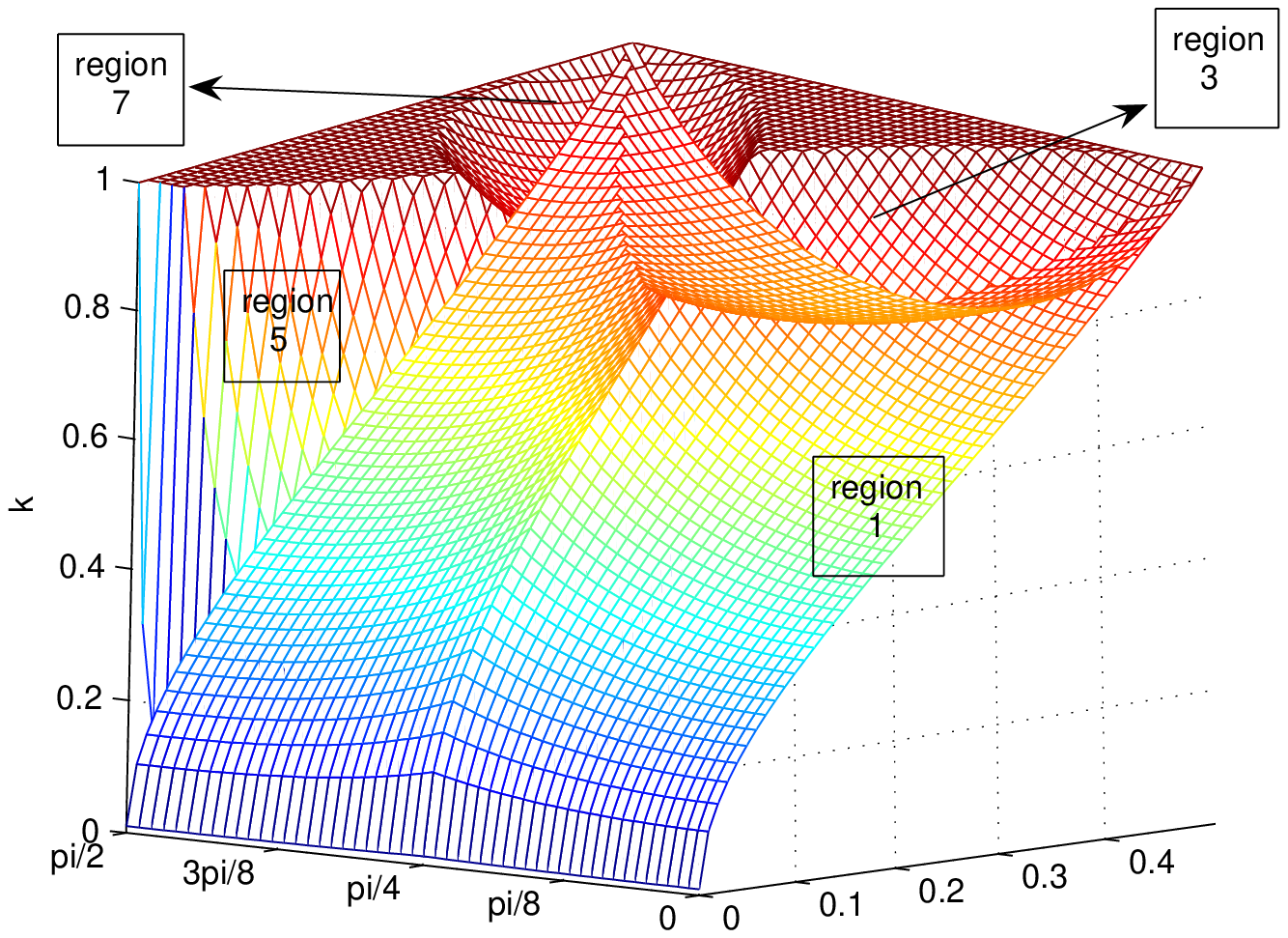}}
\subfigure[ rolling-over  view  ] {\includegraphics[height=2in,width=2in,angle=0]{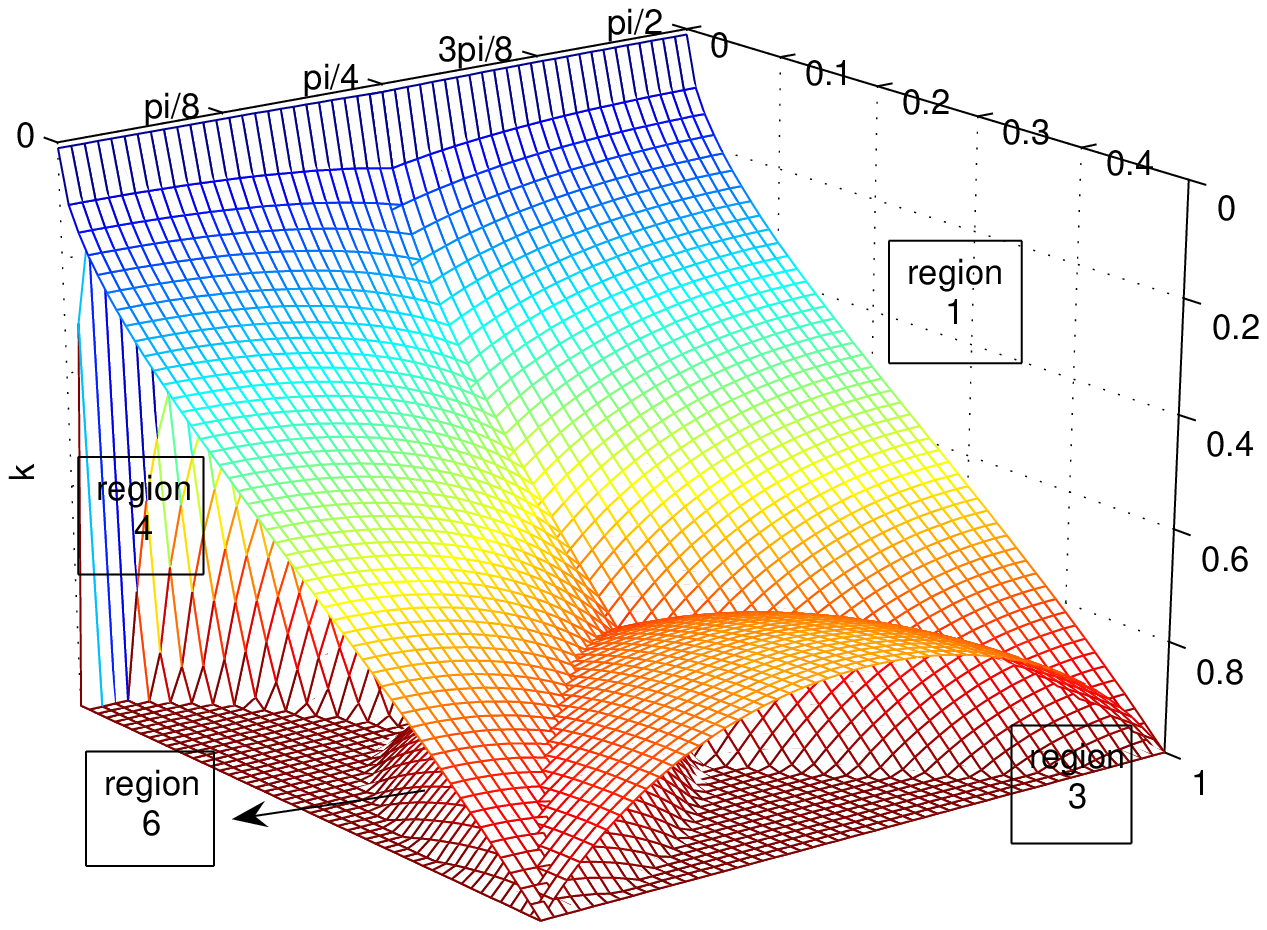}}
\caption{Complete classification on the state parameter regions.}
\label{fig5}
\end{figure*}

Fig.\ref{fig5} shows all the parameter regions of $a, k, \phi$ such that the three $B_2-$distances are achieved. These regions completely characterize all the optimal convex approximations of a sate $\rho$ w.r.t. $B_2-$distance.

It has been shown that, for a given state, the three optimal distances to the bases in $B_2^{'}$,
 $B_2^{''}$ and  $B_2^{'''}$ satisfy a ind of tradeoff relations.
In fact, the bounds on $D_{B_2}(\rho)=D_{B_2^{'}}(\rho)+D_{B_2^{''}}(\rho)+D_{B_2^{'''}}(\rho)$ or
$D_{B_2}(\rho)^2=D_{B_2^{'}}(\rho)^2+D_{B_2^{''}}(\rho)^2+D_{B_2^{'''}}(\rho)^2$ are tightly related to
the quantum uncertainty relations satisfied by the three Pauli operators, since both
the distances and the standard deviations of the Pauli operators are given by the mean values of the
the Pauli operators. From $\langle\sigma_x\rangle^2+\langle\sigma_y\rangle^2+\langle\sigma_z\rangle^2
=4k^2a(1-a)\sin^2\varphi+4k^2a(1-a)\cos^2\varphi+(1-2a)^2\leq 1$, one gets
$(\Delta S_x)^2+(\Delta S_x)^2+(\Delta S_x)^2\geq {1}/{2}$,
where $\Delta S_x~ (\Delta S_y, \Delta S_z)$ is the standard deviation and
$\Delta S_x=\frac{\sqrt{1-\langle\sigma_x\rangle^2}}{2}$.
On the other hand
$|\langle\frac{\sigma_x}{2}\rangle|+|\langle\frac{\sigma_y}{2}\rangle|+|\langle\frac{\sigma_z}{2}\rangle|\nonumber\\
=\frac{2\sqrt{2}a(1-a)+1-2a}{2} \leq \frac{3\sqrt{2}}{8}$.
Therefore, we have
$$
(\Delta S_x)^2+(\Delta S_y)^2+(\Delta S_z)^2\geq \frac{\tau}{2}(|\langle\frac{\sigma_x}{2}\rangle|+|\langle\frac{\sigma_y}{2}\rangle|+|\langle\frac{\sigma_z}{2}\rangle|),
$$
where $\tau=\frac{2}{\sqrt{3}}$ is the triple constant given in the uncertain relations in \cite{chenbin,congfeng}.
From formulae (4), (8) and (12), we immediately get that
in region $1$, our Pauli $B_2-$distances is in accordance with the uncertainty relation.

\section{Conclusion}

In summary, we have shown that a qubit mixed state $\rho$ can be approximated by a number of effectively available pure states spanned by the eigenstates of the Pauli matrices.
It is well known that correlation limits the extractable information \cite{clon,corr2,corr3,corr4}, where one does want to minimize the probability of discrimination.
The advantage of our results is that we presented the complete set of optimal decompositions of a given state. In \cite{MFSacchi} for a given state, only one particular optimal decomposition has been elegantly derived, in which $p_3$ and $p_5$ are chosen to be zero. Hence, basically it is the minimal distance with respect to four of six eigenvectors of the Pauli matrices.
As a simple example, consider the following mixed qubit state,
$\rho=\left(
\begin{array}{cc} 1/2
& 1/5\\
1/5
& 1/2\\ \end{array} \right )$.
One can verify that $D_{B_2^{'}}(\rho)=0$. All the optimal convex approximation points with respect to the basis $\{|0 \rangle ,|1\rangle ,|2\rangle ,|3\rangle \}$ are given by Eq. (\ref{pi11}). If we choose $t=0$, then we obtain $\rho=0.3\rho_0+0.3\rho_1+0.4\rho_2$. Moreover, we also have $D_{B_2^{'''}}(\rho)=0$. The optimal convex approximation points with respect to the basis $\{|2 \rangle ,|3\rangle ,|4\rangle ,|5\rangle \}$ are given by Eq. (\ref{dunoc3P3}), also for $t=0$, one obtains the optimal decomposition, $\rho=0.7\rho_2+0.3\rho_3$. In \cite{MFSacchi},
only one optimal decomposition $\rho=0.3\rho_0+0.3\rho_1+0.4\rho_3$ is obtained. Other optimal decompositions like $\rho=0.7\rho_2+0.3\rho_3$ can not be obtained even considering the optimal convex approximation with respect to the full bases $\{|0 \rangle ,|1\rangle ,|2\rangle ,|3\rangle,|4\rangle ,|5\rangle \}$.

It is obvious that $B_3-$distance $D_{B_3}(\rho)$ is always less than the $B_2-$distance $\min D_{B_{2}(\rho)}$, since the approximate point in $D_{B_2^{'}}(\rho)$, $D_{B_2^{''}}(\rho)$ and $D_{B_2^{'''}}(\rho)$ is contained in $D_{B_3}(\rho)$. For more detail, for  $k=\frac{4}{5}$,
compare $\min D_{B_{2}(\rho)}$ and  $D_{B_3}(\rho)$ in \cite{MFSacchi} in the region $a\times \phi=[0,\frac{1}{2}]\times[0,\frac{\pi}{2}]$, one can find that $\min D_{B_{2}(\rho)}=D_{B_3}(\rho)$  about twenty percent of the region, while in the remaining eighty percent region,  $D_{B_3}(\rho)$ is always less than  $\min D_{B_{2}(\rho)}$, when $a=\frac{1}{4}, \phi=\frac{\pi}{4}$, the maximal difference of $D_{B_3}(\rho)$ and $\min D_{B_{2}(\rho)}$ can be attained to 0.213, from which one can obtain that the for some case $D_{B_3}(\rho)$ is equal to  $\min D_{B_{2}(\rho)}$ while for some other case $D_{B_3}(\rho)$ is less than  $\min D_{B_{2}(\rho)}$, this is because two  eigenstates of the Pauli matrices are discarded in the computation of $B_2-$distance.
Therefore, the research of the best convex approximation about $B_2-$distance may provide an alternative way to analyze the optimal convex approximation about $B_3-$distance. Our approach may be also used to study other kinds of optimal decompositions associated with other bases.

\bigskip
\noindent {\bf Acknowledgments}
This work was completed while Bo Li was visiting the
Max-Planck-Institute for Mathematics in the Sciences in Germany under the support of the China Scholarship
Council. This work is supported by NSFC(11765016,11675113) and Jiangxi Education Department Fund (KJLD14088, GJJ161056).

\end{document}